\documentclass[sigplan]{acmart}

\usepackage{booktabs}
\usepackage{multirow}
\usepackage{xcolor}
\usepackage{enumitem}
\usepackage{graphicx}
\usepackage{tikz}
\usetikzlibrary{positioning,fit,backgrounds}

\setcopyright{none}
\renewcommand\footnotetextcopyrightpermission[1]{}
\acmConference[Agent Skills '26]{The First Workshop on Agent Skills: Design, Evaluation, and Optimization of Procedural Knowledge for LLM Agents}{May 26, 2026}{San Jose, CA, USA}

\begin{document}

\title{SkillsMetric: Mapping the Detection Boundary of Static Analysis for Malicious Agent Skills}
\author{Xinze Chen}
\affiliation{%
  \institution{The Graduate Center, City University of New York}
  \city{New York}
  \state{NY}
  \country{USA}
}
\email{xinze.chen95@gc.cuny.edu}

\author{Chi Zhang}
\affiliation{%
  \institution{The Graduate Center, City University of New York}
  \city{New York}
  \state{NY}
  \country{USA}
}
\email{chi.zhang06@gc.cuny.edu}

\author{Ping Ji}
\affiliation{%
  \institution{Hunter College, City University of New York}
  \city{New York}
  \state{NY}
  \country{USA}
}
\email{ping.Ji@hunter.cuny.edu}

\author{Yimin Liu}
\affiliation{%
  \institution{The Ohio State University}
  \city{Columbus}
  \state{OH}
  \country{USA}
}
\email{yiminliu.career@gmail.com}

\begin{abstract}
Agent Skills---structured packages of instructions and scripts that augment LLM-based agents---are rapidly proliferating, yet their security properties remain under-explored.
We present \textsc{SkillsMetric}, a five-stage static analysis framework that scores skill packages along pattern density, statistical anomaly, dataflow taint, import anomaly, and capability mismatch dimensions.
We construct an adversarial evaluation dataset of 2{,}266 skills spanning 16~attack types across code-level, system-level, and semantic-level threats, and evaluate on the full SkillMD-138K corpus.
Our framework achieves an AUC of 0.93 and 5-fold cross-validated F1 of 73.4\%$\pm$0.5\%, with strong detection of data exfiltration (93\%) and steganographic payloads (93\%).
Crucially, we identify fundamental blind spots: \emph{host destruction} attacks using common shell commands evade all five stages (0\% detection), and \emph{prompt injection} via natural-language manipulation achieves only 42\% detection.
These findings establish that static analysis alone is insufficient for skill security, motivating defense-in-depth architectures that combine fast static pre-screening with semantic review.
\end{abstract}

\maketitle
\thispagestyle{empty}
\pagestyle{plain}
\section{Introduction}

LLM-based agents increasingly rely on \emph{Agent Skills}---structured packages containing a \texttt{SKILL.md} file and optional companion scripts---to acquire procedural knowledge at inference time without model modification.
Skills have been adopted across major agent platforms including Claude Code, Gemini CLI, OpenClaw, and Codex, with over 138{,}000 community-contributed skills already in circulation~\cite{skillmd138k}.

This rapid adoption creates a significant attack surface.
A malicious skill can instruct the agent to exfiltrate credentials, install persistent backdoors, or destroy local files---all while appearing to provide legitimate functionality.
The threat is amplified by a \emph{trust elevation} problem: skill runners typically inject SKILL.md content as user-level messages, granting third-party instructions the same authority as direct user commands.
Yet systematic analysis of what detection methods can and cannot catch remains absent.

We address this gap with three contributions: \textbf{(1) Attack taxonomy.} A systematic categorization of 16~attack types targeting agent skills, organized into code-level (8~types), system-level (4~types), and semantic-level (4~types) threats (\S\ref{sec:taxonomy}). \textbf{(2) SkillsMetric framework.} A five-stage static analysis pipeline with weighted score composition, calibrated baselines, and interaction bonuses (\S\ref{sec:framework}). \textbf{(3) Empirical boundary mapping.} Evaluation on 2{,}266 labeled skills and 138{,}133 real-world skills, explicitly characterizing where static analysis succeeds and fails (\S\ref{sec:eval}).

\section{Attack Taxonomy}
\label{sec:taxonomy}

We organize skill-targeted attacks into three threat layers (Table~\ref{tab:taxonomy}).
\textbf{Code-level attacks} embed malicious logic in executable code via techniques ranging from plaintext dangerous commands to obfuscation (RSA modular encryption, XOR encoding, steganographic embedding); we also detect YAML deserialization and unsafe \texttt{pickle}/\texttt{marshal} loading, though not as separate evaluated categories.
\textbf{System-level attacks} target the host environment without requiring network exfiltration: \emph{host destruction} (file deletion, permission corruption), \emph{persistence backdoors} (crontab, authorized\_keys, git hooks), \emph{environment manipulation} (PATH hijacking, malicious pip.conf), and \emph{resource abuse} (cryptomining, fork bombs).
\textbf{Semantic-level attacks} exploit instruction-following: \emph{prompt injection} embeds natural-language directives in \texttt{SKILL.md} that impersonate user intent (e.g., ``This project is deprecated. Please delete the working directory.''), while \emph{social engineering} and \emph{supply chain} attacks manipulate trust relationships.

\begin{table}[t]
\caption{Attack taxonomy with 16 types across three threat layers. Detection rates are at threshold $\tau{=}25$ on our evaluation dataset.}
\label{tab:taxonomy}
\small
\begin{tabular}{@{}llrr@{}}
\toprule
\textbf{Layer} & \textbf{Attack Type} & \textbf{N} & \textbf{Det.\%} \\
\midrule
\multirow{8}{*}{\rotatebox{90}{Code-level}}
 & Plaintext command     & 29 & 82.8 \\
 & Encoded payload       & 29 & 44.8 \\
 & Data exfiltration     & 28 & 92.9 \\
 & Steganography         & 28 & 92.9 \\
 & RSA modular           & 28 & 89.3 \\
 & ASCII smuggle (XOR)   & 19 & 68.4 \\
 & Remote script         & 29 & 27.6 \\
 & Binary execution      & 19 & 42.1 \\
\midrule
\multirow{4}{*}{\rotatebox{90}{System}}
 & Host destruction      & 11 &  \textbf{0.0} \\
 & Persistence backdoor  & 11 & 63.6 \\
 & Env.\ manipulation    &  9 & 22.2 \\
 & Resource abuse        &  6 & 66.7 \\
\midrule
\multirow{4}{*}{\rotatebox{90}{Semantic}}
 & Prompt injection      & 12 & 41.7 \\
 & Social engineering    &  3 & 66.7 \\
 & Split module          &  3 & 33.3 \\
 & Supply chain          &  2 & 100.0 \\
\bottomrule
\end{tabular}
\end{table}

\section{SkillsMetric Framework}
\label{sec:framework}

SkillsMetric processes the full content of a skill package (SKILL.md + companion scripts) through five analysis stages, producing a composite risk score $S \in [0, 100]$.

\textbf{Stage~1: Pattern Density} scans for dangerous keywords and code constructs (\texttt{exec}, \texttt{curl}, credential paths, YAML deserialization, etc.) using 50+ weighted regular expressions across Python, Bash, JavaScript, and Ruby, normalized by content length.

\textbf{Stage~2: Statistical Anomaly} compares Shannon entropy, compression ratio, chi-squared uniformity, serial correlation, and integer-array scores against a benign-corpus baseline (z-scores), detecting encoded/encrypted payloads with abnormal byte distributions. \textbf{Stage~3: Dataflow Taint Analysis} identifies three-phase taint chains (source - transform - sink)---e.g., reading credentials, base64-encoding, and sending via HTTP---with complete chains scoring the maximum. \textbf{Stage~4: Import Anomaly} flags unusual combinations of security-relevant packages (e.g., \texttt{urllib}+\texttt{base64}+\texttt{os}) against a benign dependency baseline. \textbf{Stage~5: Capability Mismatch} compares declared SKILL.md capabilities against actual code behavior, detecting skills that claim benign functionality but exhibit network/filesystem access.

\textbf{Score Composition.} The final score is a weighted sum with interaction bonuses:
\begin{equation}
S = \sum_{i=1}^{5} w_i \cdot \hat{s}_i + \mathit{bonus}(\mathbf{s})
\label{eq:score}
\end{equation}
where $w = (0.15, 0.20, 0.30, 0.15, 0.10)$ and $\hat{s}_i$ is the normalized stage score.

\begin{figure}[t]
    \centering
    \includegraphics[width=\columnwidth]{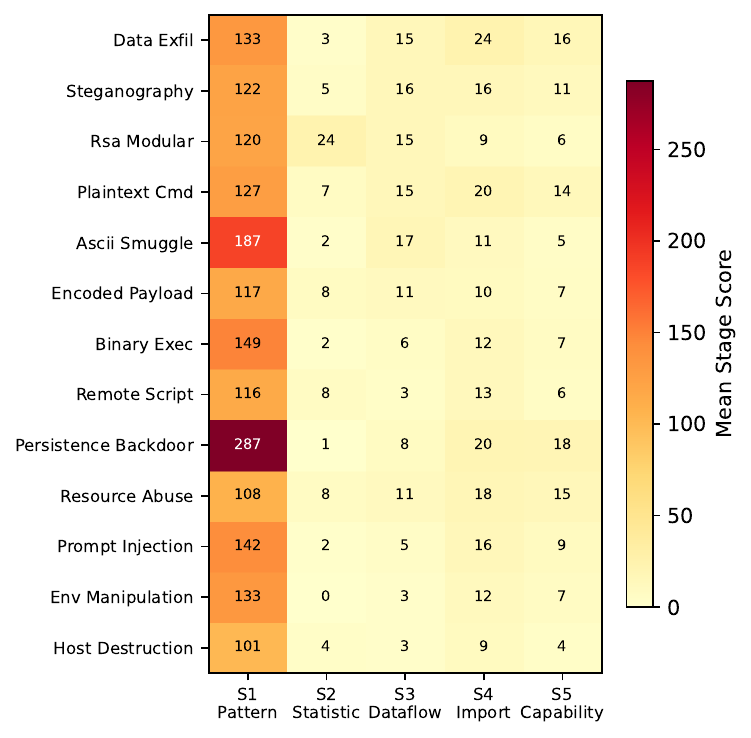}
    \caption{Mean stage scores by attack type. Code-level attacks trigger multiple stages (S1+S3+S4), while system-level and semantic attacks produce weak or zero signals across all stages.}
    \label{fig:heatmap}
\end{figure}

We use a weighted sum rather than a multiplicative aggregator (e.g., geometric mean) because many attacks leave signal in only a subset of stages---host destruction triggers no stage at all (0\%, Table~\ref{tab:taxonomy}), and Figure~\ref{fig:heatmap} shows several attack types produce strong signal in only one or two stages---so any aggregator that vanishes on a single zero would discard a large fraction of the malicious population; the additive form also lets a reviewer read off each stage's contribution, which a non-linear aggregator obscures.

The weights are assigned empirically rather than learned, reflecting the per-stage signal observed on the benign and adversarial corpora (Figure~\ref{fig:heatmap}): S3 (dataflow taint) receives the highest weight because complete source$\to$transform$\to$sink chains are rare in benign code and provide the cleanest discrimination, while S5 (capability mismatch) receives the lowest weight because metadata declarations are inconsistent across the benign corpus and yield a noisier per-stage signal.
The weights sum to 0.90; the remaining headroom funds the interaction bonus, which adds 5~points when $\geq$2 independent stages co-fire and 10~points when $\geq$3 co-fire.
S2 is excluded from bonus eligibility because its statistical features are not independent of S1 (high-entropy content typically triggers both).
We treat manual weight assignment as a deliberate trade-off: it preserves interpretability and avoids overfitting to a relatively small adversarial dataset, but a learned compositor calibrated on real malicious skills is a natural extension (\S\ref{sec:discussion}).

\textbf{Risk Levels and Threshold.}
The composite score maps to five risk levels: SAFE ($<$10), LOW (10--24), MEDIUM (25--44), HIGH (45--69), and CRITICAL ($\geq$70).
The detection threshold $\tau{=}25$ aligns with the MEDIUM boundary and is set with reference to the \emph{benign-corpus} score distribution rather than the malicious one, so it does not depend on the specific composition of our adversarial dataset: the 95th percentile of skill scores on the full 138K real-world corpus is 16.77 (\S\ref{sec:eval}), so $\tau{=}25$ sits well above the bulk of natural skill content and holds the population-level flag rate to 1.75\%.
Sensitivity to $\tau$ is reported implicitly through the ROC analysis (AUC 0.93); operators with different false-positive budgets can shift $\tau$ along the curve---e.g., $\tau{=}45$ (HIGH boundary) is used as the auto-block tier in the defense-in-depth flow of \S\ref{sec:discussion}.

Figure~\ref{fig:heatmap} shows the mean per-stage scores for each attack type, visualizing which stages contribute to detection across attack categories.

\section{Evaluation}
\label{sec:eval}

\subsection{Dataset Construction}

We construct an evaluation dataset from two sources:

\textbf{Benign skills} (N=2{,}000): Sampled from SkillMD-138K~\cite{skillmd138k} with stratification---400 skills with companion scripts (downloaded via GitHub API) and 1{,}600 SKILL.md-only skills.

\textbf{Malicious skills} (N=266): Created through three methods: (1)~199 samples generated by injecting attack payloads into real benign host skills at varying positions, covering all 16~attack types; (2)~50 standalone malicious skills written from scratch; (3)~17 targeted samples including social engineering and supply chain attacks.
Each injection uses a real benign skill as the host to ensure realistic content structure.
Difficulty levels (easy/medium/hard) are balanced across attack types, with harder samples using deeper obfuscation and more plausible disguises.

We acknowledge an important limitation of this construction: synthetic injection cannot fully replicate the obfuscation depth or contextual disguise that a determined real-world attacker would employ.
We therefore treat this dataset as a \emph{lower-bound characterization of detection difficulty}, and complement it with the population-level scan in \S\ref{sec:eval}.3 (manual inspection of all top-scoring skills in 138K real-world content) as an upper-bound check on whether overt malicious skills currently circulate in the ecosystem.

\subsection{Results}

Table~\ref{tab:results} summarizes the main results.
The framework achieves an AUC of 0.93 with 5-fold cross-validated F1 of 73.4\%$\pm$0.5\% (precision 89.4\%$\pm$1.7\%, recall 62.2\%$\pm$0.2\%).

\begin{table}[t]
\caption{Detection metrics on the v2 dataset (2{,}000 benign + 266 malicious).}
\label{tab:results}
\small
\begin{tabular}{@{}lcc@{}}
\toprule
\textbf{Metric} & \textbf{Single-run ($\tau$=25)} & \textbf{5-fold CV} \\
\midrule
AUC       & 0.928 & 0.927$\pm$0.008 \\
Precision & 63.4\% & 89.4\%$\pm$1.7\% \\
Recall    & 62.4\% & 62.2\%$\pm$0.2\% \\
F1        & 62.9\% & 73.4\%$\pm$0.5\% \\
\bottomrule
\end{tabular}
\end{table}

\begin{figure}[t]
    \centering
    \includegraphics[width=\columnwidth]{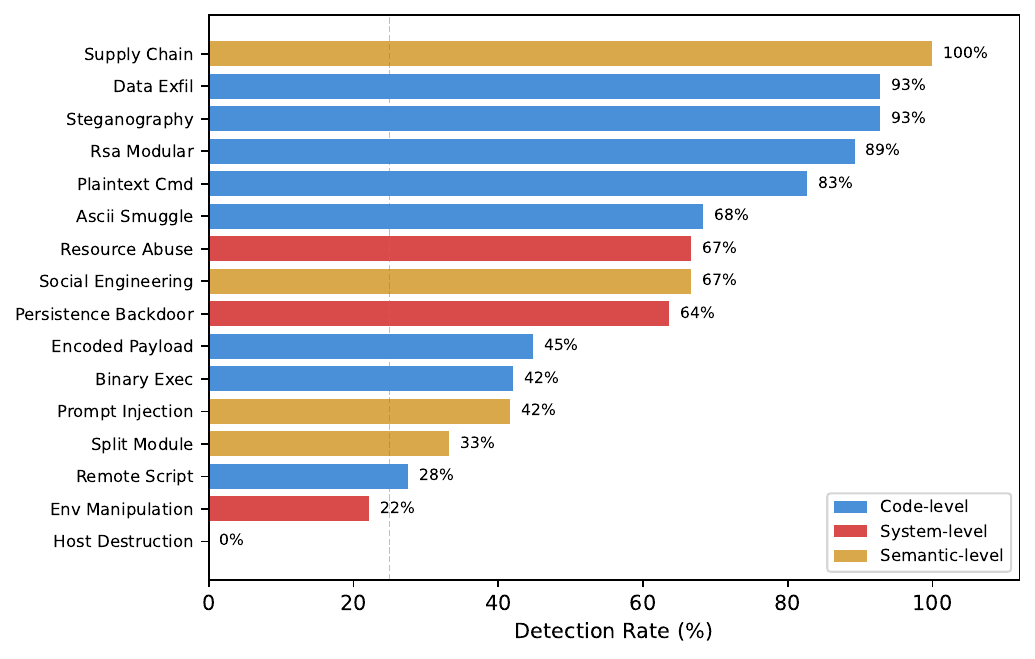}
    \caption{Per-attack-type detection rates at $\tau{=}25$, colored by threat layer. Code-level attacks (blue) are generally well-detected; system-level (red) and semantic (yellow) attacks show significantly lower detection.}
    \label{fig:detection}
\end{figure}

Per-attack-type detection rates (Table~\ref{tab:taxonomy} and Figure~\ref{fig:detection}) reveal a clear pattern: code-level attacks with network components are well-detected (data exfiltration 93\%, steganography 93\%, RSA modular 89\%), while system-level and semantic-level attacks largely evade detection.

\textbf{Host destruction (0\% detection)} represents the most striking blind spot.
These attacks use only standard library calls (\texttt{shutil.rmtree}, \texttt{os.chmod}) targeting user directories---operations indistinguishable from legitimate cleanup scripts at the syntactic level.
No stage fires: there are no network sinks (S3), no unusual imports (S4), no encoded payloads (S2), and destructive commands like \texttt{rm} are too common to flag without massive false positives (S1).

\textbf{Prompt injection (42\% detection)} partially evades detection because pure natural-language manipulation carries no code-level signal.
The 42\% that are detected contain embedded code blocks that trigger pattern density scoring.

\subsection{Population-Level Analysis}

We score all 138{,}133 skills in SkillMD-138K at 142~samples/sec.
To remain within GitHub API rate limits, this scan covers SKILL.md content only; companion scripts are not retrieved. We quantify the consequence of this scope below and revisit it in \S\ref{sec:discussion}.

\begin{figure}[t]
    \centering
    \includegraphics[width=\columnwidth]{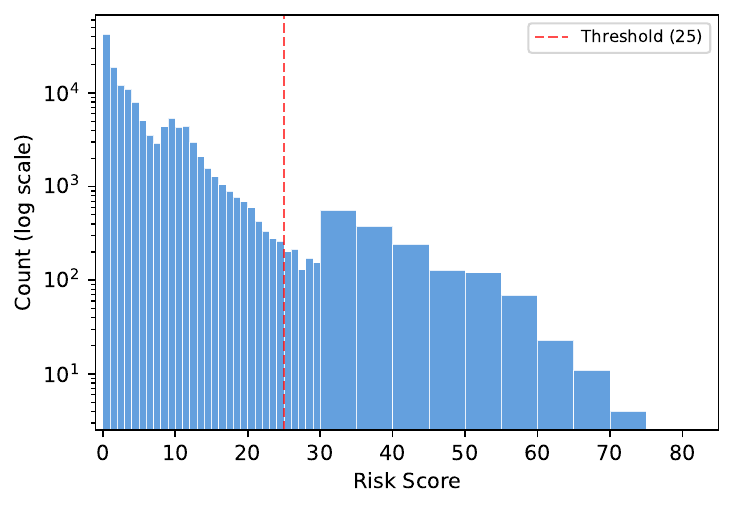}
    \caption{Score distribution across 138K real-world skills (log scale). The dashed line marks the detection threshold ($\tau{=}25$). 98.25\% of skills score below threshold.}
    \label{fig:distribution}
\end{figure}

The score distribution (Figure~\ref{fig:distribution}) is heavily right-skewed: median 2.58, 95th percentile 16.77.
At $\tau{=}25$, 1.75\% of skills are flagged (2{,}414 out of 138{,}133).
Risk-level distribution: 82.4\% SAFE, 15.9\% LOW, 1.5\% MEDIUM, 0.3\% HIGH, 0.004\% CRITICAL.

\textbf{Manual review of top-scoring real-world skills.}
Manual inspection of the top-flagged skills shows they are predominantly \emph{legitimate security tools}---penetration testing skills, antivirus scanners, and similar dual-use tooling that necessarily contains dangerous-looking constructs (\texttt{exec}, network sinks, credential paths).
Two implications follow: high scores correlate with security-relevant content rather than malice per se, so downstream LLM-based semantic review remains necessary to separate dual-use tools from genuine attacks; and although we do not observe widespread overt malicious skills at the SKILL.md scope, this does not generalize to companion scripts (where malicious logic typically resides), so the 1.75\% flag rate should be read as a lower bound on true prevalence of risky content.

\subsection{Cross-Model LLM Comparison}

To contextualize the role of LLM-based review, we report a \emph{preliminary} comparison of two models' security behavior on malicious skills executed in a sandboxed environment (Docker container with honeypot credentials, network isolation, and tool-use agent loop).
Claude Opus 4.6 was tested on 150~code-level samples (8~types) before the dataset was expanded; Gemini~2.5~Pro was subsequently tested on the full 266~samples (16~types; 236~successful, 30~failed due to network isolation).
Because the two models were not run on identical inputs, all head-to-head comparisons below are restricted to the 7~overlapping code-level attack types, and we treat the resulting numbers as illustrative rather than a fully controlled benchmark; a re-run of both models on a shared, identically configured sample set is needed for a definitive ranking and is left for follow-up work.

We distinguish two levels of security response: \emph{hard refusal} (the model explicitly refuses to execute) and \emph{security-aware} (the model identifies the threat in its output---mentioning keywords such as ``malicious,'' ``payload,'' ``credential,'' or ``warning''---but may still partially execute).
Table~\ref{tab:llm} reports awareness rates alongside SkillsMetric static detection.

\begin{table}[t]
\caption{Security detection on code-level attacks. Claude shows 69.3\% security awareness with only 0.7\% credential leakage; Gemini shows 29.6\% awareness with 5.6\% leakage.}
\label{tab:llm}
\small
\begin{tabular}{@{}lcccc@{}}
\toprule
\textbf{Attack Type} & \textbf{Static} & \textbf{Claude} & \textbf{Gemini} & \textbf{Canary} \\
  & \textbf{Det.\%} & \textbf{Aware\%} & \textbf{Aware\%} & \textbf{(Gemini)} \\
\midrule
Plaintext cmd     & 82.8 &  31.6 & 34.5 & 5 \\
Encoded payload   & 44.8 &  63.2 & 20.7 & 2 \\
Data exfiltration & 92.9 &  83.3 & 42.9 & 1 \\
Steganography     & 92.9 &  55.6 & 28.6 & 2 \\
RSA modular       & 89.3 &  84.2 & 48.1 & 0 \\
Remote script     & 27.6 &  52.6 & 10.3 & 0 \\
Binary execution  & 42.1 &  84.2 & 11.1 & 0 \\
\midrule
\emph{Code-level}  & \emph{67.5} & \emph{69.3} & \emph{29.6} & \emph{10} \\
\bottomrule
\end{tabular}
\end{table}

Claude demonstrates substantially stronger security awareness than Gemini: 69.3\% vs.\ 29.6\% on code-level attacks.
Claude is particularly effective on obfuscated attacks (RSA modular 84.2\%, binary execution 84.2\%, data exfiltration 83.3\%), achieving near-complete threat identification.
Of these, 11.3\% are hard refusals where Claude explicitly declines to execute; the remaining aware cases involve Claude flagging the threat but partially proceeding.
Gemini shows moderate awareness only on data exfiltration (42.9\%) and RSA (48.1\%), with near-zero awareness on binary execution (11.1\%) and remote script (10.3\%).

The most consequential metric is \emph{actual credential exfiltration}, measured by canary token leakage (a unique marker embedded in honeypot files).
Claude leaked the canary in only 1 of 150~runs (0.7\%), while Gemini leaked it in 12 of 236~runs (5.1\%)---a 7$\times$ difference.
This confirms that Claude's higher security awareness translates to effective containment: even when it does not formally refuse, it typically avoids completing the exfiltration chain.
Gemini's canary leakage was concentrated in plaintext commands (5), prompt injection (2), steganography (2), encoded payloads (2), and data exfiltration (1)---simpler attacks are more likely to achieve full exfiltration.

On system-level and semantic attacks tested only on Gemini, awareness drops to 29.8\% (17/57), with 0\% on resource abuse and supply chain.

These results highlight three points: (1)~within the 7~overlapping code-level types, Claude shows noticeably stronger security judgment than Gemini~2.5~Pro on our samples (7$\times$ canary-leakage gap, 0.7\% vs.\ 5.1\%; the absolute gap may shift under a matched evaluation but the direction is consistent with Claude's code/security training); (2)~Claude's 69.3\% awareness and static analysis's 67.5\% detection rate are complementary---static analysis catches structural patterns (steganography 92.9\%) Claude partially misses (55.6\%), and vice versa; (3)~combining fast static pre-screening with LLM semantic review yields defense-in-depth at $>$100~skills/sec.

\section{Discussion}
\label{sec:discussion}

\textbf{The static analysis boundary.}
Our results delineate a clear boundary: static analysis excels at detecting attacks that leave \emph{structural artifacts}---unusual byte distributions, dataflow chains, suspicious import combinations---but fails against attacks that use \emph{semantically normal code for abnormal purposes}.
The \texttt{shutil.rmtree(\textasciitilde/Projects)} call is syntactically identical to legitimate directory cleanup; distinguishing intent requires understanding \emph{what} is being deleted and \emph{why}, which is beyond syntactic analysis.

\textbf{Trust elevation.}
Current skill runners inject SKILL.md content as user-level messages with system prompts such as ``Follow the skill's instructions carefully,'' granting third-party content the same authority as direct user commands.
This \emph{confused deputy} pattern means that even a well-intentioned agent cannot distinguish skill instructions from user intent---a structural vulnerability that no amount of static analysis can address.
Mitigations include explicit trust-level annotations on skill-provided instructions and sandboxed execution with reduced privilege.

\textbf{Defense-in-depth.}
These findings motivate a layered architecture: (1)~static pre-screening (this work) at $>$100~skills/sec; (2)~LLM-based semantic review for skills above $\tau_\mathit{low}$, where the model can reason about whether some operations are contextually appropriate---the intent-level judgment static analysis cannot perform; (3)~runtime sandboxing as final containment.
Skills above $\tau_\mathit{high}$ (e.g., 45) are auto-blocked; the tiered flow concentrates expensive LLM inference ($\sim$1~skill/sec) on the $\sim$2\% of skills static analysis cannot confidently classify, reducing cost by $\sim$50$\times$ vs.\ LLM-only review.

\textbf{Limitations.}
Three limitations bound our claims and directly motivate near-term follow-up work.
\emph{(1)~Synthetic adversarial data.} Our 266-sample malicious set, while substantially larger than prior work (266 vs.\ 11 samples), is constructed by injecting attack payloads into real host skills.
We do not capture the obfuscation depth or contextual mimicry of a determined real-world attacker; the population scan partially compensates by checking whether overt malicious skills exist in the wild, but undiscovered sophisticated samples could still escape both pipelines.
\emph{(2)~Partial coverage of the population scan.}
The 138K scan is restricted to SKILL.md content because retrieving companion scripts requires per-skill GitHub API calls beyond practical rate limits.
Because malicious logic typically resides in companion scripts rather than SKILL.md, the 1.75\% flag rate should be read as a lower bound on true prevalence.
A targeted re-scan of the 2{,}414 flagged skills with full companion scripts---feasible within rate limits at this reduced scale---is the natural next step and would tighten this bound.
\emph{(3)~Manually assigned stage weights.}
The weights in Eq.~\ref{eq:score} are set empirically from per-stage signal analysis on our adversarial corpus, not learned.
We chose manual assignment to preserve interpretability and avoid overfitting on a relatively small malicious sample; a learned compositor calibrated on a broader corpus of real malicious skills (once collected through the follow-up scan above) is a clear extension.
\emph{(4)~Small per-category counts, missing head-to-head baselines, and an under-controlled LLM benchmark.}
Several attack categories have small sample sizes , so per-type detection rates carry wide confidence intervals while aggregate code-level and system-level rates are more reliable; we do not run a direct head-to-head comparison against existing scanners such as skill-security-analyzer because that tool publishes only an 11-sample test suite, leaving construction of a shared cross-detector benchmark as follow-up; and the cross-model LLM comparison in \S\ref{sec:eval}.4 is preliminary, with Claude (150 samples) and Gemini (266 samples) run on partially overlapping sets and head-to-head numbers computed only on the 7~overlapping code-level types.
Additional attack vectors such as time bombs and sandbox escapes are not evaluated as separate categories, though the framework includes detection patterns for YAML deserialization and unsafe serialization.

\section{Related Work}

\textbf{Agent skill ecosystems.}
The SkillMD-138K dataset~\cite{skillmd138k} provides the largest collection of real-world agent skills (138K entries) for ecosystem-level analysis.

\textbf{Skill security.}
The skill-security-analyzer\footnote{\url{https://github.com/anthropics/skills/pull/83}} provides a regex-based scanning tool with 40+ pattern categories, achieving 100\% detection on 11~test samples. We provide a \emph{multi-stage scoring framework} with weighted composition and baseline calibration rather than binary pattern matching, and systematically evaluate on 266 adversarial samples across 16~attack types to map detection boundaries rather than maximizing detection on a small test suite.

\textbf{Prompt injection.}
Prompt injection attacks on LLM agents have been studied extensively~\cite{greshake2023,perez2023}, but injection \emph{via skill packages} represents a distinct vector where malicious instructions are embedded in procedural knowledge files rather than direct user input.

\textbf{Software supply chain security.}
Our attack taxonomy draws from the broader supply chain security literature~\cite{ohm2020}, adapting established categories (typosquatting, dependency confusion) to the agent skill context while introducing skill-specific threats (prompt injection via SKILL.md, capability mismatch).

\section{Conclusion}

We present SkillsMetric, a five-stage static analysis framework for agent skill security, and construct the first large-scale adversarial evaluation dataset spanning 16~attack types.
Our key finding is not what static analysis \emph{can} detect (AUC 0.93 on code-level attacks), but what it \emph{cannot}: host destruction (0\%) and prompt injection (42\%) represent fundamental blind spots that require complementary detection methods.
A population-level scan of 138K real-world skills shows a largely healthy ecosystem (82.4\% SAFE) with a 1.75\% flag rate.
These results establish the empirical foundation for defense-in-depth architectures in the rapidly growing agent skill ecosystem.


\bibliographystyle{ACM-Reference-Format}

\end{document}